\documentclass[12pt]{article}
\pdfoutput=1
\usepackage[pdftex]{graphicx}
\usepackage{amssymb}
\usepackage{amsmath}
\usepackage{bm}
\usepackage{braket}
\usepackage{comment}
\usepackage{setspace}
\usepackage{geometry}
\usepackage{color}
\usepackage[colorlinks=true,linkcolor=blue,citecolor=blue]{hyperref}

\begin{document}
\begin{titlepage}

\begin{flushright}
ICRR-Report-653-2013-2\\
IPMU 13-0104
\end{flushright}

\vskip 1.35cm

\begin{center}
{\Large {\sf Domain wall and isocurvature perturbation problems \\
in axion models}}

\vskip 1.3cm

{Masahiro Kawasaki$^{a,b}$, Tsutomu T. Yanagida$^b$ and Kazuyoshi Yoshino$^a$} \\

\vskip 0.7cm

{\it$^a$Institute for Cosmic Ray Research, University of Tokyo, Kashiwa, Chiba 277-8582, Japan}\\
\vskip 0.3cm
{\it $^b$Kavli Institute for the Physics and Mathematics of the Universe(WPI), 
Todai Institutes for Advanced Study,, University of Tokyo, Kashiwa, Chiba 277-8583, Japan}\\
\end{center}

\vskip 1.3cm
\begin{abstract}
Axion models have two serious cosmological problems, domain wall 
and isocurvature perturbation problems. 
In order to solve these problems we investigate the Linde's model 
in which 
the field value of the Peccei-Quinn (PQ) scalar is large 
during inflation. 
In this model the fluctuations of the PQ field grow after inflation through the parametric
resonance and stable axionic strings may be produced, 
which results in the domain wall problem. 
We study formation of axionic strings 
using lattice simulations. 
It is found that in chaotic inflation the axion model is free from both the domain wall and 
the isocurvature perturbation problems if the initial misalignment 
angle $\theta_{a}$ is smaller than $O(10^{-2})$.
Furthermore, axions can also account for the dark matter 
for the breaking scale $v\simeq10^{12-16}\,\mathrm{GeV}$ and 
the Hubble parameter during inflation $H_{\mathrm{inf}}\lesssim 10^{11-12}\,\mathrm{GeV}$ 
in general inflation models.

\end{abstract}
\end{titlepage}

\tableofcontents

\section{Introduction}
\label{sec:Introduction}

Axion~\cite{Weinberg:1977ma;Wilczek:1977pj} is a scalar particle predicted 
in Peccei-Quinn (PQ) mechanism~\cite{Peccei:1977ur;Peccei:1977hh} which is a natural 
solution to the strong CP problem in QCD. 
In the PQ mechanism there exits a compelex scalar field with global $U(1)_{\mathrm{PQ}}$ symmetry.
The $U(1)_{\mathrm{PQ}}$ symmetry is spontaneously broken at some scale $v$
and axion is a Nambu-Goldstone boson associated with it. 
The axion is also attractive in cosmology because the coherent oscillation 
of the axion field behaves like the nonrelativistic fluid and  
can be a good candidate for the dark matter 
of the universe~\cite{Preskill:1982cy;Abbott:1982af;Dine:1982ah}. 

In the cosmological scenario, however, the axion models cause the domain wall 
problem~\cite{Sikivie:1982qv}. 
When the cosmic temperature falls to the symmetry breaking scale $v$, 
$U(1)_{\mathrm{PQ}}$ symmetry is spontaneously broken, which leads to formation of 
one-dimensional topological defects called axionic strings.  
Furthermore, when the cosmic temperature cools down as low as the QCD scale, the axion potential 
is lifted up through the QCD instanton effect and the axion acquires its mass.
Since the axion potential has $N_{\rm DW}$ ($N_{\rm DW}$ is called domain wall number) 
desecrate minima and the axion field settles down to one of the minima,
domain walls are formed so that $N_{\rm DW}$ domain walls attach each string. 
$N_{\rm DW}$ is determined by QCD anomaly and it depends on details of axion models. 
For example, $N_{\mathrm{DW}}$ is the number of heavy quarks which have $U(1)_{\mathrm{PQ}}$ 
charge in the KSVZ model~\cite{Kim:1979if;Shifman:1979if}, 
while it is double of the number of generations, i.e. $N_{\rm DW}=6$, 
in the DFSZ model~\cite{Dine:1981rt;Zhitnitsky:1980tq}. 
It is known that stable domain walls are disastrous in cosmology because they
dominate the universe soon after their formation 
and overclose the universe, which contradicts the present observations.

There are a few solutions to the domain wall problem. 
One is the $N_{\mathrm{DW}}=1$ model which can be realized in the KSVZ model. 
The domain walls with $N_{\mathrm{DW}}=1$ are disk-like objects whose boundaries are strings
and they collapse by their tension~\cite{Vilenkin:1982ks}.
Thus, string-wall networks with $N_{\mathrm{DW}}=1$ are unstable. 
In this case, domain walls decay into axion particles soon after their formation. 
In Ref.~\cite{Hiramatsu:2012gg} the spectrum of axions radiated from strings and domain walls 
is calculated numerically and it was found that the axion decay constant $F_{a}$, 
which is related with the PQ breaking scale as $v=N_{\mathrm{DW}}F_{a}$, should satisfy 
$F_{a}\lesssim(2.0-3.8)\times10^{10}\,\mathrm{GeV}$ in order that the axion abundance 
does not exceed the dark matter abundance. 
Another solution for $N_{\mathrm{DW}}\ge 2$ is to postulate the bias parameter 
which breaks the PQ symmetry explicitly~\cite{Sikivie:1982qv}. 
Due to the bias, the degeneracy of the potential minimum is resolved and domain walls 
become unstable. 
However, this bias also shifts the minimum of the axion potential and hence violates CP.
In Ref.~\cite{Hiramatsu:2012sc} it was found that the phase of the bias parameter 
needs to be fine-tuned in order to satisfy the constraint from CP violation as well as 
cosmological and astrophysical constraints.

The above arguments apply to the case where the PQ symmetry breaking occurs after inflation. 
Assuming that the PQ symmetry is spontaneously broken before or during inflation, 
the domain wall problem is expected to be solved for the general domain number $N_{\mathrm{DW}}$ 
because the exponential cosmic expansion during inflation makes the value of the axion field 
homogeneous in the whole observable universe.\footnote{
At least, it is necessary that $U(1)_{\mathrm{PQ}}$ breaking scale is larger than 
the reheating temperature in order that the $U(1)_{\mathrm{PQ}}$ symmetry is not restored 
thermally after inflation.}
However, the axion field obtains fluctuation $\delta a=H_{\mathrm{inf}}/2\pi$
with $H_{\mathrm{inf}}$ being the Hubble parameter during inflation, 
which leads to large isocurvature density 
perturbations~\cite{Axenides:1983hj,Seckel:1985tj,Linde:1985yf,Linde:1990yj,Turner:1990uz,Lyth:1991ub}.
The isocurvature density perturbations are stringently constrained by cosmic microwave
background (CMB) observations~\cite{Hinshaw:2012aka,Ade:2013uln}.  
Therefore, when the PQ symmetry breaking takes place during or before inflation, we
have another cosmological difficulty, i.e., isocurvature perturbation problem. 
In particular, this problem is serious for the chaotic inflation model~\cite{Linde:1983gd}
because of its large Hubble parameter during inflation and the axion model may be 
inconsistent with chaotic inflation.\footnote{
Moreover, the domain wall problem may recur in the chaotic inflation model
since the phase of the PQ complex field has the random value due to the large fluctuations 
of the axion field during inflation. }
 
In Ref.~\cite{Linde:1991km}, Linde proposed that the large expectation value in the radial 
direction of the PQ field during inflation can suppress the isocurvature perturbations 
and it can avoid the domain wall problem simultaneously. 
After inflation, though, the large fluctuations of the PQ field are generated by the parametric 
resonance~\cite{Kofman:1995fi,Kofman:1997yn,Shtanov:1994ce} and they can lead to
nonthermal restoration of the $U(1)_{\mathrm{PQ}}$ 
symmetry~\cite{Tkachev:1995md,Tkachev:1998dc,Kasuya:1997ha,Kasuya:1998td}. 
Then, stable axionic strings are formed at the subsequent 
$U(1)_{\mathrm{PQ}}$ symmetry breaking 
and eventually the domain wall problem comes again after the QCD phase transition
~\cite{Kasuya:1996ns,Kasuya:1997td}. 
The formation of the stable strings with the nonthermal symmetry 
restoration was calculated using lattice simulations in the radiation dominated background 
after chaotic inflation with quartic potential in Ref.~\cite{Kasuya:1999hy}. 
It was found that the $U(1)_{\mathrm{PQ}}$ symmetry breaking scale must satisfy 
$v\gtrsim3\times10^{16}\,\mathrm{GeV}$ in order that the stable axionic strings 
which lead to the domain wall problem are not formed.\footnote{
This constraint depends on the initial value of the PQ field at the beginning of oscillation. 
In Ref.~\cite{Kasuya:1999hy} it is presumed to be the Planck scale $M_{\mathrm{p}}$.} 

In this paper, we reexamine the formation of the axionic strings in Linde's model 
using lattice simulation. 
We assume that the universe is matter dominated after inflation in contrast 
to Ref.~\cite{Kasuya:1999hy}, which is natural when the inflaton oscillates along the quadratic 
potential.
We find that the constraint on the PQ breaking scale is much relaxed as 
$v\gtrsim 10^{-4}\left|\Phi\right|_{i}$,
where $\left|\Phi\right|_{i}$ represents the initial value of the PQ field. 
Together with observational constrains, it is found that chaotic inflation
is consistent with the axion model if the initial misalignment 
angle $\theta_{a}$ is less than $O(10^{-2})$. 
Furthermore, we find that axion can be dark matter without the isocurvature perturbation  
problem nor the domain wall problem  
for $v\simeq10^{12-16}\,\mathrm{GeV}$ and $H_{\mathrm{inf}}\lesssim 10^{11-12}\,\mathrm{GeV}$ 
in general inflation models.

This paper is organized as follows. 
Section~\ref{sec:The Dynamics of the Fields} introduces our model 
and studies the analytical prediction for the constraint of $U(1)_{\mathrm{PQ}}$ 
breaking scale. 
In Section~\ref{sec:Numerical Simulations}, the numerical simulation 
for the nonlinear dynamics after inflation is carried out. 
Section~\ref{sec:Observational Constraints} describes the observational constraints 
on the model parameters. 
We summaries our conclusion in Section~\ref{sec:Conclusion}.

\section{Dynamics of the Fields}
\label{sec:The Dynamics of the Fields}

Let us consider an inflaton field $\chi$ and a complex PQ field $\Phi$ with the potential
\begin{equation}
   V=\frac{M^2}{2}\chi^2+\frac{\lambda}{2}\left(\left|\Phi\right|^2-v^2\right)^2,
   \label{eq:potential}
\end{equation}
where $M$ is the mass of the inflaton, $\lambda$ is a self-coupling constant of the PQ field 
and $v$ is the breaking scale of $U(1)_{\mathrm{PQ}}$ symmetry. 
Here, for concreteness we consider the chaotic inflation model 
and the inflaton has a value larger than Planck scale during inflation. 
We assume that the value in the radial direction of the PQ field during inflation 
is large enough to satisfy the observational constraint on isocurvature perturbations. 
When the PQ field satisfies the slow-roll condition, it follows the attractor solution 
given by~\cite{Harigaya:2012up}
\begin{equation}
  \left|\Phi\right|=\frac{M}{\sqrt{2\lambda}}\left(\ln\frac{\chi_{0}}{\chi}\right)^{-1/2},
  \label{eq:attractor solution}
\end{equation}
where $\chi_0$ is the inflaton value when the inflaton escapes 
from the stochastic region, given by
\begin{equation}
   \chi_{0}\sim3\times10^{3}M_{\mathrm{p}}\left(\frac{10^{13}\,\mathrm{GeV}}{M}\right)^{1/2}.
   \label{eq:initial inflaton}
\end{equation}
In the chaotic inflation model, the field value of the inflaton corresponding to 
e-folding number $N=60$ is $\chi_{N=60}\simeq15M_{\mathrm{p}}$ and the inflation ends at 
$\chi_{\mathrm{end}}=\sqrt{2}M_{\mathrm{p}}$. 
Therefore, from Eqs.~(\ref{eq:attractor solution}) and (\ref{eq:initial inflaton}) we obtain 
$\left|\Phi\right|_{N=60}\simeq1.2\left|\Phi\right|_{\mathrm{end}}$. 
Namely, the PQ field hardly move during inflation.

After inflation the universe is dominated by the oscillation of the inflaton. 
Since the effective mass of the PQ field in the radial direction is 
$m_{\mathrm{eff}}\simeq\sqrt{3\lambda}\left|\Phi\right|\simeq0.8H$ at the end of inflation, 
the PQ field starts to oscillate soon after inflation. 
When the amplitude of the PQ field is much larger than $v$, the potentail of the PQ field 
is approximately quartic and the amplitude of the PQ field oscillation decreases 
as $a^{-1}$. 
Thus, the scale factor $a_{c}$ when the PQ field settles down to the minimum 
of its potential is estimated as 
\begin{equation}
   a_{c}\simeq\frac{\left|\Phi\right|_{i}}{v}a_{i},
   \label{eq:critical time}
\end{equation}
where the subscript $``\, i \, "$ represents the initial time at the beginning of oscillation. 
Hereafter we adopt the normalization as $a_{i}=1$.

Until the homogeneous mode of the PQ field settles down to the potential minimum, 
the fluctuations of the PQ field grow exponentially through the parametric resonance. 
If the amplitude of the fluctuations $\delta\Phi$ become larger than $v$, 
the effective potential is lifted up and the PQ symmetry is restored nonthermally. 
After that, the fluctuations decrease by the cosmic expansion, $U(1)_{\mathrm{PQ}}$ symmetry 
is spontaneously broken again and stable cosmic strings which lead to the domain wall 
formation may eventually be formed. 

We perform lattice simulations and examine whether stable cosmic strings are formed or not. 
Before showing the results of the numerical simulations, let us estimate 
the order of the PQ scale $v$ for strings not to be formed. 
We define two real scalar fields $X=\mathrm{Re}\,\Phi$ and $Y=\mathrm{Im}\,\Phi$, 
and these two fields can be decomposed into their homogeneous parts and fluctuations 
as $X=\bar{X}+\delta X$ and $Y=\bar{Y}+\delta Y$ with the initial conditioins 
$\bar{X}_{i}=\left|\Phi\right|_{i}, \bar{Y}_{i}=0$. 
When the amplitude of $\bar{X}$ is much larger than $v$, the time evolution of $\bar{X}$ is approximately given by~\cite{Greene:1997fu}
\begin{equation}
   \bar{X}(\tau)\simeq\frac{\left|\Phi\right|_{i}}{a(\tau)}
   \cos\left(c\sqrt{\lambda}\left|\Phi  \right|_{i}\left(\tau-\tau_{\mathrm{i}}\right)\right),
   \label{eq:homogeneous solution}
\end{equation}
where $\tau=\int dt/a$ is the conformal time and $c\simeq0.8472$ is a constant.\footnote{
In the matter dominated universe the eqaution of motion contains $\frac{a''}{a}=\frac{2}{\tau^2}$ 
unlike the radiation dominated universe in~\cite{Greene:1997fu}. 
We neglect this term in the analytical estimation for simplicity.}
The linearized equations of motion of the fluctuations in Fourier space are
\begin{eqnarray}
   \ddot{\delta X}_{k}+3H\dot{\delta X}_{k}-\frac{k^2}{a^2}\delta X_{k}
   +\lambda\left(3\bar{X}^2-v^2\right)\delta X_{k}&=&0,
   \label{eq:real fluctuation eom}\\
   \ddot{\delta Y}_{k}+3H\dot{\delta Y}_{k}-\frac{k^2}{a^2}\delta Y_{k}
   +\lambda\left(\bar{X}^2-v^2\right)\delta Y_{k}&=&0.
   \label{eq:imaginary fluctuation eom}
\end{eqnarray}
Rescaling as $z=c\sqrt{\lambda}\left|\Phi\right|_{i}\left(\tau-\tau_{i}\right)$, 
$\delta x_{k}=a\delta X_{k}$, $\delta y_{k}=a\delta Y_{k}$, 
Eqs.\,\eqref{eq:real fluctuation eom} and \eqref{eq:imaginary fluctuation eom} 
become Mathieu equation:
\begin{eqnarray}
   \frac{d^2\delta x_{k}}{dz^2}+\left[A_{x}+2q_{x}\cos2z\right]\delta x_{k}&\simeq&0,
   \label{eq:real Mathieu}\\
   \frac{d^2\delta y_{k}}{dz^2}+\left[A_{y}+2q_{y}\cos2z\right]\delta y_{k}&\simeq&0,
   \label{eq:imaginary Mathieu}
\end{eqnarray}
where
\begin{equation}
   A_{x}=\frac{k^2-a^2\lambda v^2}{c^2\lambda\left|\Phi\right|_{i}^2}+2q_{x},\ \ 
   A_{y}=\frac{k^2-a^2\lambda v^2}{c^2\lambda\left|\Phi\right|_{i}^2}+2q_{y},\ \ 
   q_{x}=\frac{3}{4c^2},\ \ 
   q_{y}=\frac{1}{4c^2}.\nonumber
\end{equation}
Here we have neglected the term which contains the decreasing factor 
$\frac{a''}{a}=\frac{2}{\tau^2}$ for simplicity. 
From Eq.\,\eqref{eq:critical time} and $q_{x}\simeq1.04, q_{y}\simeq0.35$, 
$A_{x}$ is larger than unity for any momentum, 
on the other hand $A_{y}$ can be unity for an appropriate momentum. 
Therefore the parametric resonance occurs in the first instability band for $\delta y_{k}$, 
but it occurs in the second instability band for $\delta x_{k}$. 
Since the resonance of $\delta y_{k}$ is stronger than that of $\delta x_{k}$, 
the growth of the fluctuations of the imaginary part of the PQ field is faster than 
that of the real part.
This is also understood from the fact that there is no potential in the imaginary direction.
Neglecting the fluctuations of the real part $\delta X$ for the above reason, 
the amplitude of the field fluctuations is estimated as 
\begin{eqnarray}
  \braket{\left|\delta \Phi\right|^2}&\simeq&\braket{\delta Y^2}\nonumber\\
   &=&\frac{1}{a^2}\int\frac{dk}{k}\frac{k^3}{2\pi^2}\left|\delta y_{k}\right|^2.
  \label{eq:variance of fluctuation1}
\end{eqnarray}
Assuming that the initial condition for the fluctuations is the flat spectrum 
$\frac{k^3}{2\pi^2}\left(\frac{H_{\mathrm{inf}}}{2\pi}\right)^2$ which is generated 
during inflation, and that the first instability band dominates the momentum integration, 
we have
\begin{eqnarray}
   \braket{\left|\delta \Phi\right|^2} 
   & \simeq &\frac{1}{a^2}\left(\frac{H_{\mathrm{inf}}}{2\pi}\right)^2
   \frac{\Delta k_{*}}{k_{*}}e^{2\mu_{k_{*}}z}\nonumber\\
   & = &\frac{1}{a^2}\left(\frac{H_{\mathrm{inf}}}{2\pi}\right)^2
   \frac{1}{4c\sqrt{c^2-\frac{1}{2}}}
   e^{\frac{\sqrt{\lambda}\left|\Phi\right|_{\mathrm{i}}}{4c}\left(\tau-\tau_{i}\right)},
   \label{eq:variance of fluctuation2}
\end{eqnarray}
where the typical momentum and the width in the first instability band are 
$k_{*}\simeq\sqrt{\lambda}\left|\Phi\right|_{i}\sqrt{c^2-\frac{1}{2}}$ 
and $\Delta k_{*}\simeq\frac{\sqrt{\lambda}\left|\Phi\right|_{i}}{4c}$, 
the growth rate of $\delta y_{k}$ in the first instability band is 
$\mu_{k_{*}}=1/(8c^2)$~\cite{Shtanov:1994ce}. 
Moreover, the initial conformal time is 
$\tau_{i}=\frac{2}{H_{i}}\simeq\frac{2}{\sqrt{3\lambda}\left|\Phi\right|_{i}}$ 
and the conformal time when the homogeneous mode settles down 
into the minimum of its potential 
is $\tau_{c}=\sqrt{\Phi_{i}/v}\,\tau_{i}$ from Eq. \eqref{eq:critical time} 
since $a\propto\tau^2$ and $aH\tau=2$ in the matter dominated universe. 
From Eq. \eqref{eq:variance of fluctuation2}, the condition for strings not to be formed 
$\braket{\left|\delta\Phi\right|^2}(\tau_{c})\lesssim v^2$ is given by
\begin{equation}
   v\gtrsim\left|\Phi\right|_{i}\left[1+2\sqrt{3}c\ln\left(16\pi^2c\sqrt{c^2-\frac{1}{2}}
   \left(\frac{\left|\Phi\right|_{i}}{H_{\mathrm{inf}}}\right)^2\right)\right]^{-2}.
\label{eq:analytic condition}
\end{equation}
This expression is independent of the self-coupling constant $\lambda$, 
because the duration of resonance is dependent on only the ratio 
$\left|\Phi\right|_{i}/v$ from Eq. \eqref{eq:critical time} 
and the strength of resonance $\sqrt{\lambda}\left|\Phi\right|/m_{\mathrm{eff}}$ 
is constant. 
Since the Hubble parameter during the chaotic inflation is 
$H_{\mathrm{inf}}\simeq10^{13}\,\mathrm{GeV}$, the above condition is 
$v\gtrsim3\times10^{14}\,\mathrm{GeV}$ for $\left|\Phi\right|_{i}=M_{\mathrm{p}}$, 
$v\gtrsim5\times10^{13}\,\mathrm{GeV}$ for $\left|\Phi\right|_{i}=0.1M_{\mathrm{p}}$ 
and $v\gtrsim7\times10^{12}\,\mathrm{GeV}$ for $\left|\Phi\right|_{i}=0.01M_{\mathrm{p}}$. 
The results of the numerical simulations in the next section show that 
Eq.~(\ref{eq:analytic condition}) slightly overestimates 
the condition because the back reaction is not taken into account. 

The condition~(\ref{eq:analytic condition}) and the result obtained in the next section is much
weaker than that given in~\cite{Kasuya:1999hy} where the universe is radiation dominated 
after inflation.
Since the scale factor $a$ is proportional to $t^{1/2}$ in the radiation dominated universe,
the PQ field oscillates $\sim (|\Phi|_i/v)$ times until it settles down to the potential
minimum, which should be compared with $\sim (|\Phi|_i/v)^{1/2}$ 
in the case of the matter dominated universe. 
Thus, the parametric resonance is more significant for the case considered 
in~\cite{Kasuya:1999hy} 
and the more strong condition is obtained.

\section{Numerical Simulations}
\label{sec:Numerical Simulations}

In order to study the precise evolution of the PQ field after inflation, 
we have performed the lattice simulation in two dimensions.\footnote{
When we consider the growth of the fluctuation due to the parametric resonance 
after inflation, it is shown that the results of the lattice simulations 
in two and three dimensions are not different from each other~\cite{Kasuya:1999hy}. 
We have confirmed it with our code.} 
From Eq.\,\eqref{eq:potential} the equation of motion of the PQ field is as follows:
\begin{equation}
   \ddot{\Phi}+3H\dot{\Phi}-\frac{\nabla^2}{a^2}\Phi
   +\lambda\left(\left|\Phi\right|^2-v^2\right)\Phi=0.
   \label{eq:eom}
\end{equation}
Let us rescale the variables as
\begin{equation}
   d\bar{\tau}=\sqrt{\lambda}\left|\Phi\right|_{i}\frac{dt}{a},\ \ 
   d\bar{x}=\sqrt{\lambda}\left|\Phi\right|_{i}dx,\ \ 
   \varphi=a\frac{\Phi}{\left|\Phi\right|_{i}}.
   \label{eq:rescaling}
\end{equation}
%
Then, the initial rescaled conformal time is $\bar{\tau}_{i}=2/\sqrt{3}$ 
and Eq.\,\eqref{eq:eom} is written as
\begin{equation}
   \varphi''-\frac{2}{\bar{\tau}^2}\varphi-\nabla^2_{\bar{x}}\varphi
   +\left[
      \left|\varphi\right|^2
      -a^2\left(\frac{v}{\left|\Phi\right|_{i}}\right)^2
   \right]\varphi=0,
   \label{eq:dimensionless eom}
\end{equation}
where the prime represents the derivative with respect to $\bar{\tau}$ 
and we have used the time dependence of the scale factor $a\propto\tau^2$ 
in the matter dominated universe. 
We solve numerically Eq.\,\eqref{eq:dimensionless eom} on a $1024^2$ lattice
with the 4th-order symplectic integrator~\cite{Yoshida:1990zz,Hiramatsu:2010yn}. 
Since our interest is whether cosmic strings are formed or not,
we choose box size which is larger than horizon size 
and lattice size which is smaller than string width 
at any time of the simulations 
for each set of parameters. 
Furthermore, we choose total number of time steps 
so that the parametric resonance ends before the final time of every simulation. 
Table \ref{table:simulation parameters} shows some examples of simulation parameters.
The initial condition of the fluctuations of the PQ field 
$\delta\Phi/\Phi_{i}$
is taken as random numbers whose amplitude is in the range between $0$ and 
$H_{\mathrm{inf}}/(2\pi\left|\Phi\right|_{i})$ 
because variance of the PQ field fluctuations during inflation is given by 
$H^2_{\mathrm{inf}}/(2\pi)^2$.
The Hubble parameter during inflation is fixed to $H_{\mathrm{inf}}=10^{13}\,\mathrm{GeV}$
as is required for the chaotic inflation model. 
In order to check the accuracy of the code, we calculate the total energy density 
of this system without cosmic expansion and confirm that the energy is conserved with
error less than $0.1\%$. 

\begin{table}[htdp]
	\caption{Some examples of rescaled comoving box size, rescaled conformal time and resolution of lattice
	at the end of simulations for some $v$ and $\left|\Phi\right|_{i}$. 
	$\delta_{\mathrm{st}}(=1/\sqrt{\lambda}v)$ represents the width of a string.}
\begin{center}
\begin{tabular}{cc|ccc}
\hline\hline
 $v\,(\mathrm{GeV})$ & $\left|\Phi\right|_{i}$ & $\bar{L}$ & $\bar{\tau}_{f}-\bar{\tau}_{i}$ & $\left.{\delta_{\mathrm{st}}\big/dx}\right|_{f}$\\
 \hline
 $7\times10^{13}$ & $M_{\mathrm{p}}$ & \\
 $7\times10^{12}$ & $0.1M_{\mathrm{p}}$ & $350$ & $320$ & $1.3$\\
 $7\times10^{11}$ & $0.01M_{\mathrm{p}}$ & \\
 \hline
 $2\times10^{14}$ & $M_{\mathrm{p}}$ \\
 $2\times10^{13}$ & $0.1M_{\mathrm{p}}$ & $250$ & $200$ & $1.6$\\
 $2\times10^{12}$ & $0.01M_{\mathrm{p}}$ \\
 \hline
 $7\times10^{14}$ & $M_{\mathrm{p}}$ \\
 $7\times10^{13}$ & $0.1M_{\mathrm{p}}$ & $200$ & $140$ & $1.2$\\
 $7\times10^{12}$ & $0.01M_{\mathrm{p}}$ \\
 \hline\hline
\end{tabular}
\end{center}
\label{table:simulation parameters}
\end{table}

\begin{figure}[tbp]
\begin{center}
   \includegraphics[width=10cm, clip]{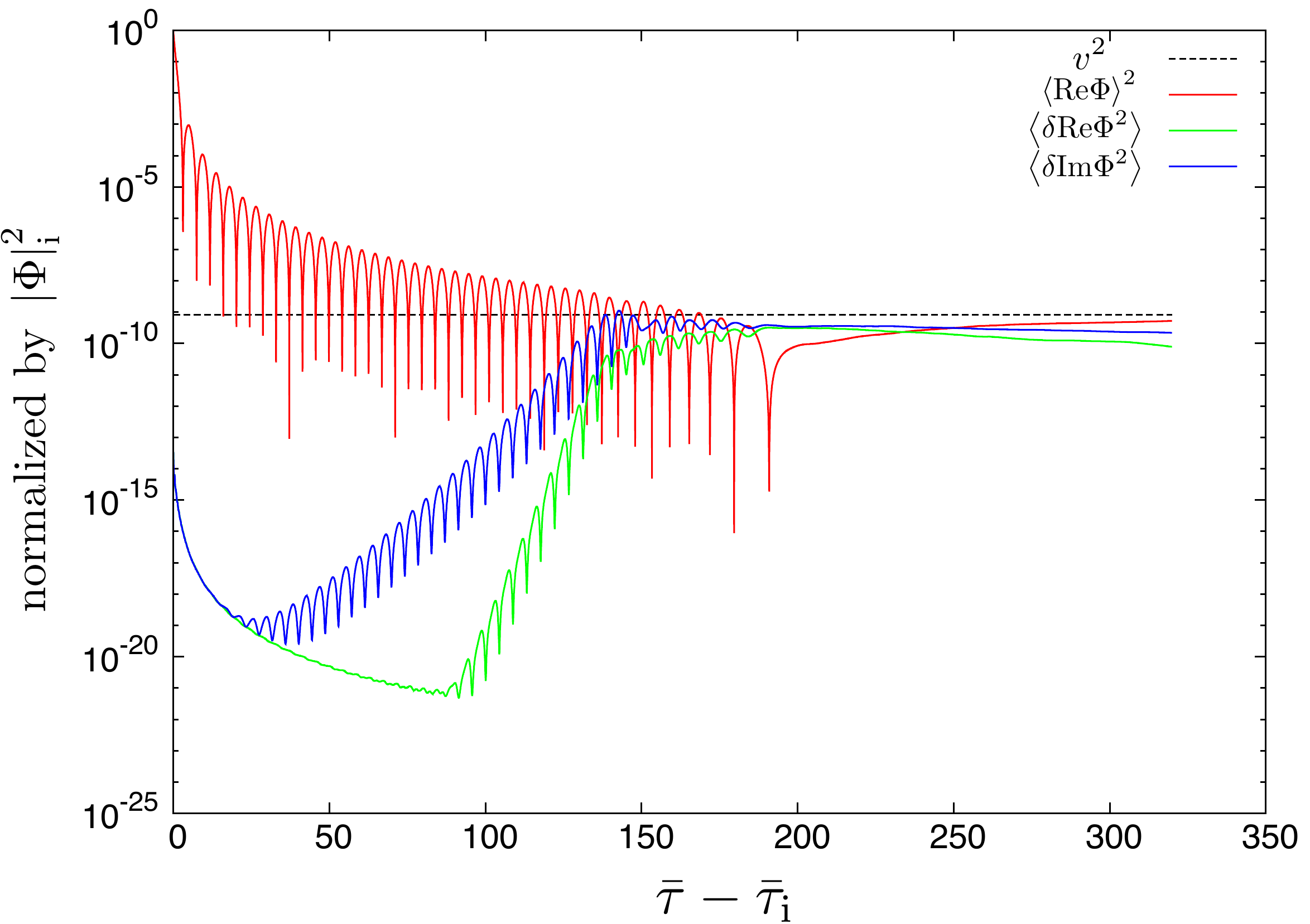}
   \caption{Time evolution of the square of the homogeneous real part of PQ field (red line) 
   and the variances of the real part (green line) and the imaginary part (blue line) 
   for $\left|\Phi\right|_{\mathrm{i}}=M_{\mathrm{p}}, v=7\times10^{13}\,\mathrm{GeV}$.
   The dashed black line denotes the square of PQ breaking scale.  
   The vertical axis is normalized by the square of the initial amplitude of the homogeneous part 
   $\left|\Phi\right|^2_{i}$.}
\label{fig:time evolution}
\end{center}
\end{figure}

\begin{figure}[tbp]
\begin{center}
\includegraphics[width=10cm, clip]{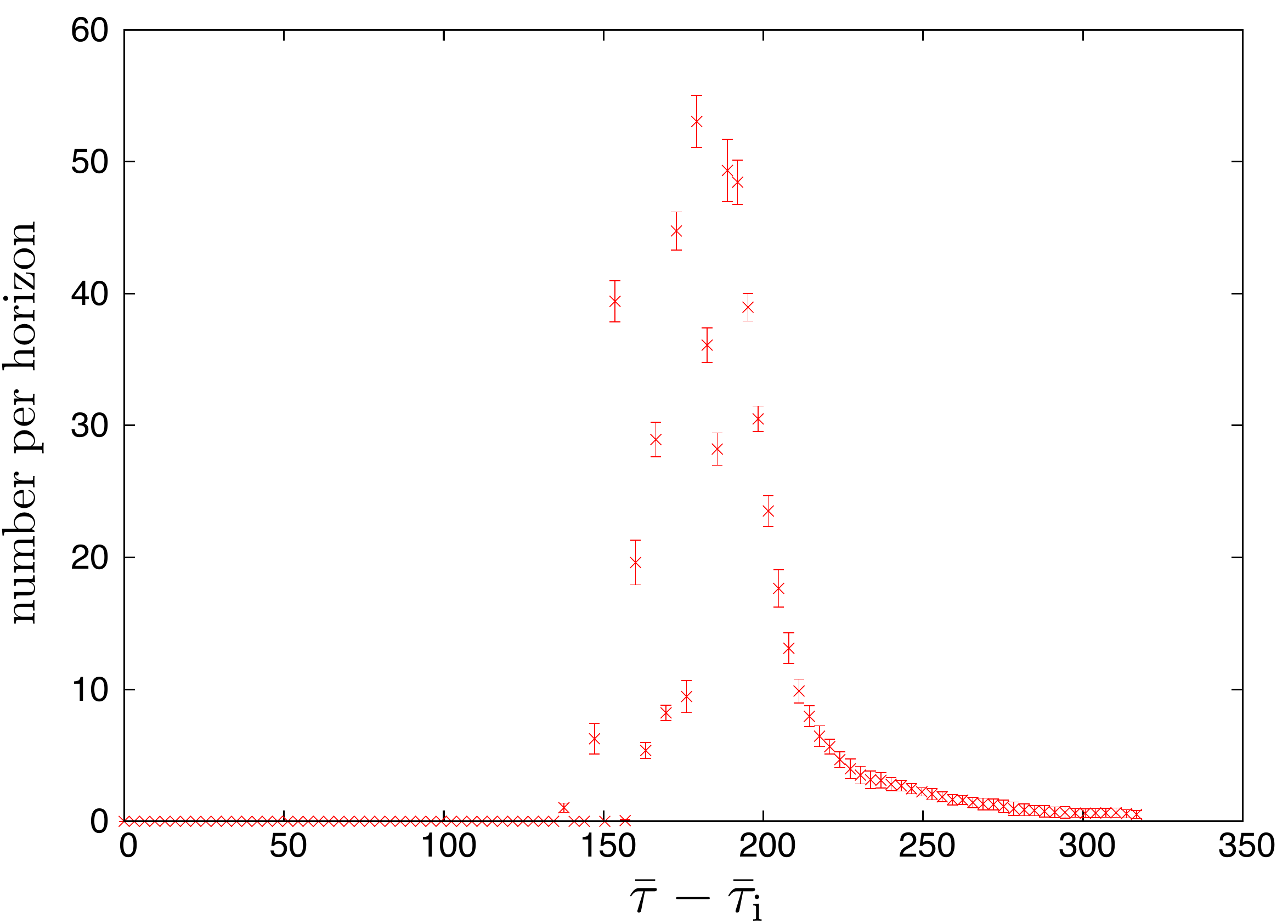}
    \caption{Average number of strings per horizon as the function of conformal time 
    for $\left|\Phi   \right|_{i}=M_{\mathrm{p}}$ and $v=7\times10^{13}~\mathrm{GeV}$. 
    We perform the simulation for $10$ realizations.}
   \label{fig:number of string}
\end{center}
\end{figure}

Varying the breaking scale $v$ of $U(1)_{\mathrm{PQ}}$, we perform the simulations 
for three initial values; $\left|\Phi\right|_{i}=M_{\mathrm{p}}$, $0.1M_{\mathrm{p}}$
and $0.01M_{\mathrm{p}}$. 
Fig.~\ref{fig:time evolution} shows a example of the evolution of the homogenous part 
and the variance of the fluctuation of the PQ field for 
$\left|\Phi\right|_{i}=M_{\mathrm{p}}$ and $v=7\times10^{13}~\mathrm{GeV}$. 
We define the real part as the direction of the initial value $\Phi_{i}$. 
As mentioned before, the fluctuations of the imaginary part grows faster than 
those of the real part because the imaginary part does not have the potential. 
In Fig.\,\ref{fig:time evolution}, it is found that the amplitude of the fluctuations
becomes larger than the breaking scale and the homogeneous part oscillates around the origin 
with amplitude smaller than the beraking scale $v$. 
Therefore, the $U(1)_{\mathrm{PQ}}$ symmetry is restored nonthermally and cosmic strings 
are formed. 
The number of strings per horizon for these model parameters is shown 
in Fig.\,\ref{fig:number of string}. 
Here, we identify the cosmic strings with the method in~\cite{Hiramatsu:2010yu}. 
We can find that the exponential growth of the fluctuations is followed by the turbulent stage 
when many strings are formed, and that the number of strings per horizon remains $O(1)$ 
after the homogeneous part settles down. 

When we search the model parameters for which strings are not formed, 
there is one subtle problem. 
In some simulations, strings are temporally formed and disappear soon 
after the homogeneous part settles, which does not lead to formation of the domain walls. 
Thus, in this paper, adopting  the criterion in~\cite{Kasuya:1999hy}
we judge stable strings are formed if strings per horizon remains almost constant. 
From this criterion, we find that the condition for the stable strings not to be formed is 
$v\gtrsim2\times10^{14}~\mathrm{GeV}$ for $\left|\Phi\right|_{i}=M_{\mathrm{p}}$, 
$v\gtrsim2\times10^{13}~\mathrm{GeV}$ for $\left|\Phi\right|_{i}=0.1M_{\mathrm{p}}$ and
$v\gtrsim2\times10^{12}~\mathrm{GeV}$ for $\left|\Phi\right|_{\mathrm{i}}=0.01M_{\mathrm{p}}$. 
Performing some simulations with larger grid points $(2048^2)$
 in order to check these results,
we confirmed that they are not changed.
Therefore, we can derive the following constraint on the breaking scale:
\begin{equation}
   v \gtrsim 1\times10^{-4}\left|\Phi\right|_{i}.
   \label{eq:constraint for v}
\end{equation}
Namely, the growth of the fluctuations of the PQ field depends on only the duration of 
the homogeneous oscillation determined by the ratio $\left|\Phi\right|_{i}/v$, 
which is consistent with the analytic estimation~(\ref{eq:analytic condition}).

\section{Observational Constraints}
\label{sec:Observational Constraints}

In this section, we consider the observational constraints on the present axion model. 
The first constraint comes from the cosmic density of axions. 
Since the coherent oscillation of the axion field after the QCD phase transition behaves like the nonrelativistic  fluid and gives a significant contribution to the dark matter density.
Thus, the axion density should satisfy 
$\Omega_{a}h^2\le\Omega_{\mathrm{CDM}}h^2=0.12$~\cite{Ade:2013lta}.

When axionic strings and domain walls are not formed after inflation, 
the axion abundance in the present universe is given 
by~\cite{Turner:1985si,Bae:2008ue,Wantz:2009it}\footnote{
We assume that there is no significant entropy production after the beginning 
of axion oscillation, that the effective degrees of freedom of radiation energy 
at the beginning of coherent oscillation is $g_{*}=61.75$ and that QCD energy scale 
is $\Lambda_{\mathrm{QCD}}=400\,\mathrm{MeV}$. Moreover, the anharmonic effect can 
be neglected because we are considering the small initial misalignment angle.}
\begin{equation}
    \Omega_{a}h^2 \simeq 0.18
    \left(\theta^2_{a}+\braket{\delta\theta^2_{a}}\right)
    \left(\frac{F_{a}}{10^{12}\,\mathrm{GeV}}\right)^{1.19},
    \label{eq:axion abundance}
\end{equation}
where $F_{a}$ is the axion decay constant, $\theta_{a}$ is the background initial 
misalignment angle defined by the axion field value at the beginning of coherent oscillation 
$a_{1}=F_{a}\theta_{a}$ and $\braket{\delta\theta^2_{a}}$ is the spatial dispersion 
of fluctuations generated during inflation. 
In our model, the radial direction of the PQ field has a large expectation value 
during inflation. 
Thus, the fluctuations of misalignment angle can be suppressed as
\begin{equation}
   \braket{\delta\theta^2_{a}} \simeq 
   \frac{N_{\mathrm{DW}}^2}{\left|\braket{\Phi}\right|^2}
   \left(\frac{H_{\mathrm{inf}}}{2\pi}\right)^2,
   \label{eq:dispersion}
\end{equation}
where we use the relation between the misalignment angle $\theta_{a}$ and the phase 
of the PQ field $\theta$, $\theta_{a}=N_{\mathrm{DW}}\theta$. 

If axionic strings are formed after inflation, 
the above expressions are not valid since the misalignment angle has large inhomogeneity in space. 
Thus, we need to replace $\theta^2_{a}+\braket{\delta\theta^2_{a}}$ with averaged value 
$2\times (\pi^2/3)$, where we take the anharmonic effect into account~\cite{Turner:1985si}. 
In case of $N_{\mathrm{DW}}\ge2$, the string-wall system is stable and it leads to 
the domain wall problem. 
On the other hand, the domain wall problem  can be avoided for $N_{\mathrm{DW}}=1$
since the string-wall system decays soon after formation. 
In this cace, strings and domain walls decay into axion particles which contribute 
to the cosmic axion density. 
In fact, this additional contribution dominates over 
that from the coherent oscillation and the abundance in the present universe is
estimated as~\cite{Hiramatsu:2012gg} 
\begin{equation}
   \Omega_{a}h^2\simeq8.8\left(\frac{F_{a}}{10^{12}\,\mathrm{GeV}}\right)^{1.19}.
   \label{eq:axion abundance from string-wall}
\end{equation}
%

The second constraint is imposed from CMB observations of the CDM isocurvature perterbations. 
The fluctuations of the axion field during inflation produce 
the CDM isocurvature perturbations whose power spectrum is 
given by~\cite{Kawasaki:2008sn,Hikage:2012be}
\begin{equation}
   \mathcal{P}_{\mathcal{S}_{\mathrm{CDM}}}(k) \simeq 
   4\left(\frac{\Omega_{a}}{\Omega_{\mathrm{CDM}}}\right)^2
   \frac{\mathcal{P}_{\delta\theta_{a}}(k)}{\theta^2_{a}+\braket{\delta\theta^2_{a}}},
   \label{eq:CDM isocurvature}
\end{equation}
where $\mathcal{P}_{\delta\theta_{a}}(k)$ is the power spectrum of $\delta\theta_{a}$. 
$\mathcal{P}_{\mathcal{S}_{\mathrm{CDM}}}$ is constrained from the last CMB observation 
as~\cite{Ade:2013uln}
\begin{equation}
   \beta_{\mathrm{iso}} \equiv 
   \frac{\mathcal{P}_{\mathcal{S}_{\mathrm{CDM}}}(k_{0})}
   {\mathcal{P}_{\zeta}(k_{0})+\mathcal{P}_{\mathcal{S}_{\mathrm{CDM}}}(k_{0})}
   <0.036 ~~~~~\mathrm{at}~~95\%~\mathrm{CL},
\end{equation}
where $\mathcal{P}_{\zeta}(k_{0})$ is the power spectrum of the curvature perturbations and
$k_{0}=0.002\,\mathrm{Mpc}^{-1}$. 
This leads to the constraint on the model parameters via Eq.~(\ref{eq:CDM isocurvature}).
If strings and domain walls are formed after inflation, 
perturbations of the misalignment angle generated during inflation
disappear due to the restoration of PQ symmetry.
Therefore, there is no constraint from the observation of CDM isocurvature perturbations.

%
%

\subsection{Chaotic inflation model}
\label{subsec:chaotic}

Now we apply the observational constraints to our model and examine the allowed 
parameter region.  
First we assume the chaotic inflation model with the quadratic potential 
and take the Hubble parameter during inflation to be the large value 
$H_{\mathrm{inf}}\simeq10^{13}~\mathrm{GeV}$.
The cosmological effect of domain walls is quite different  between 
$N_{\mathrm{DW}}=1$ and $N_{\mathrm{DW}}\ge 2$, so we discuss two cases 
separately.

\subsubsection{$N_{\mathrm{DW}}\ge2$}
\label{subsec:N>=2}

\begin{figure}[tbp]
\begin{tabular}{cc}
  \begin{minipage}{0.5\hsize}
  \begin{center}
     \includegraphics[width=7cm, clip]{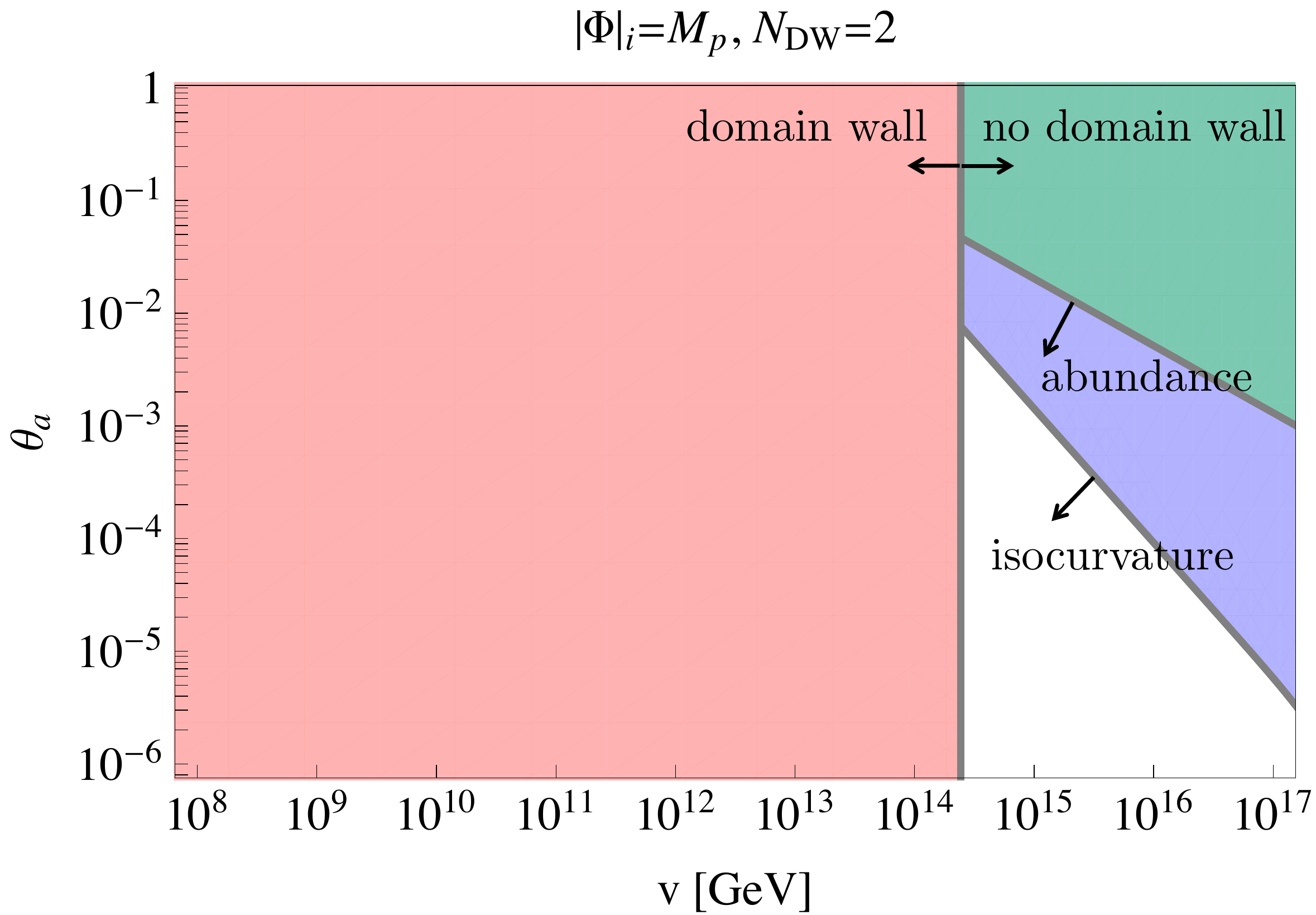}
  \end{center}
  \end{minipage}
  \begin{minipage}{0.5\hsize}
  \begin{center}
     \includegraphics[width=7cm, clip]{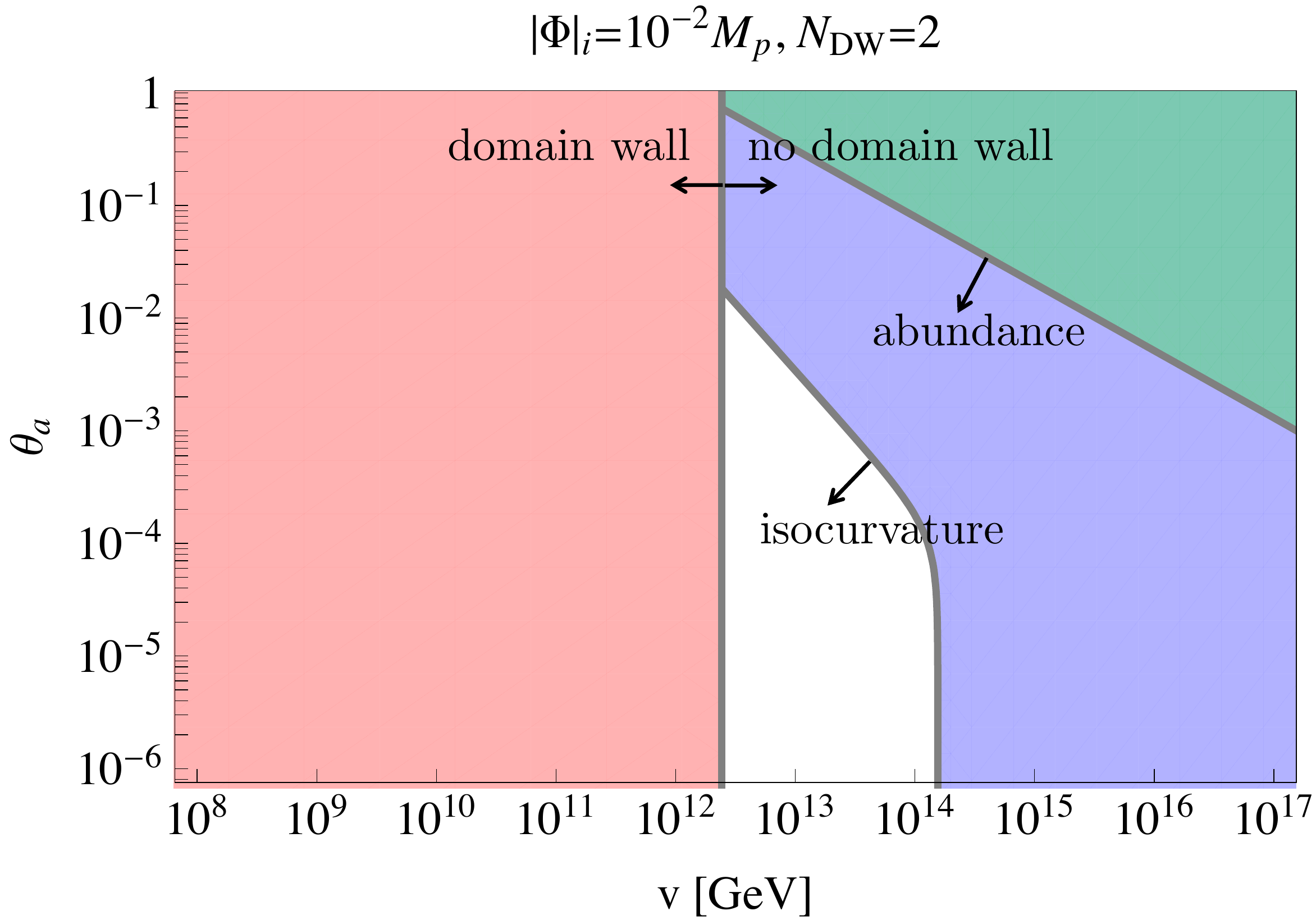}
  \end{center}
  \end{minipage}
\end{tabular}
  \begin{center}
    \includegraphics[width=7cm, clip]{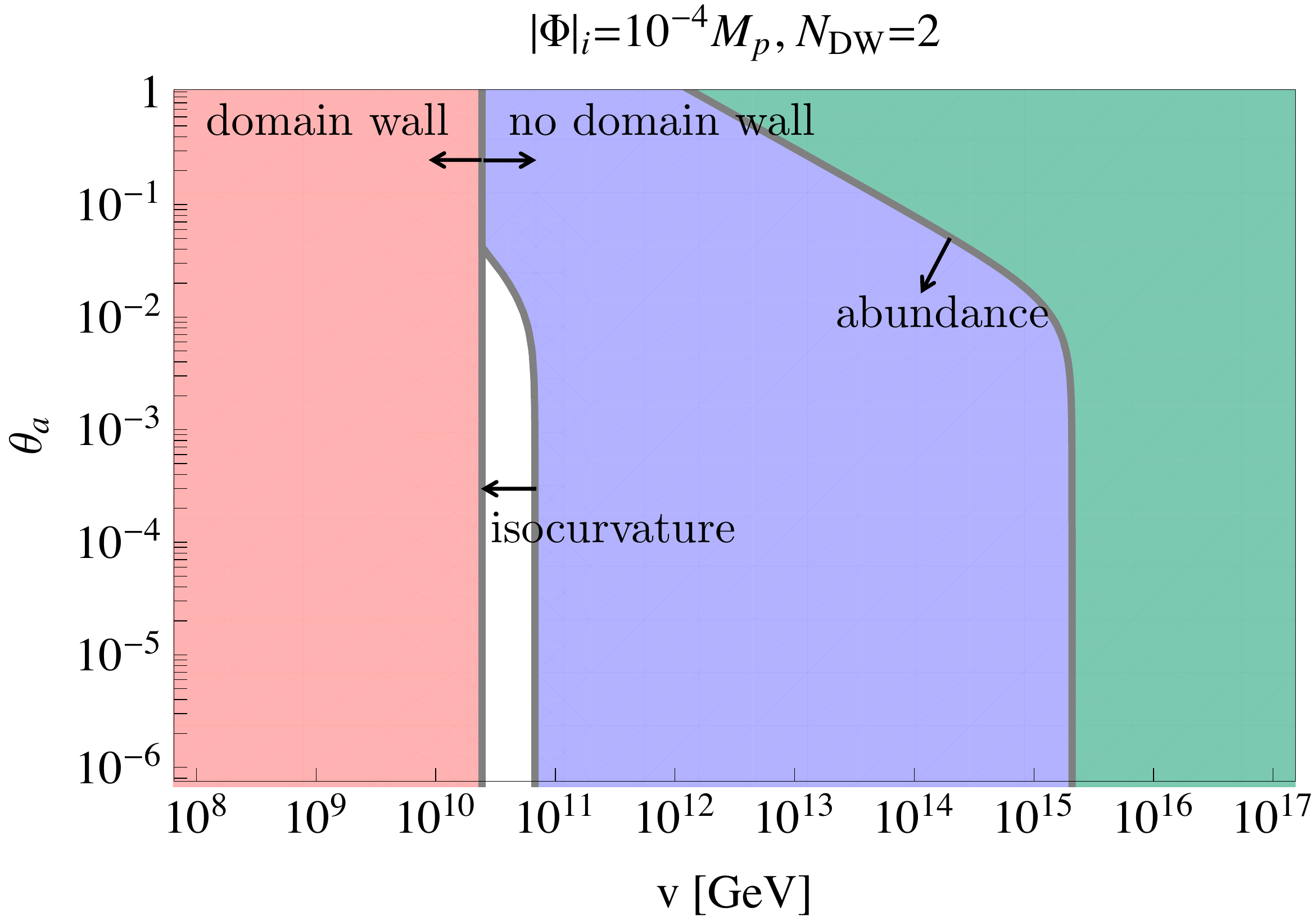}
  \end{center}
    \caption{Constraints on $v~(=N_{\mathrm{DW}}F_{a})$ and $\theta_{a}$ for $N_{\mathrm{DW}}=2$ and 
    various values of initial amplitude $\left|\Phi\right|_{i}$ in the chaotic inflation model, i.e.~$H_{\mathrm{inf}}=10^{13}\,\mathrm{GeV}$. 
    The red region is excluded in order to avoid the domain wall problem from our numerical 
    simulation. 
    The blue region and the green region are excluded by the observation of isocurvature 
    perturbations and dark matter abundance, respectively.}
   \label{fig:v-theta N=2}
\end{figure}

In case of $N_{\mathrm{DW}}\ge2$, strings and domain walls are stable once they are formed. 
Therefore, from our numerical simulations the $U(1)_{\mathrm{PQ}}$ breaking scale must satisfy 
$v\gtrsim1\times10^{-4}\left|\Phi\right|_{i}$ in order to avoid the domain wall problem. 
Fig.\,\ref{fig:v-theta N=2} shows the result of our simulation and the observational constraints 
for $N_{\mathrm{DW}}=2$ and three initial conditions 
$\left|\Phi\right|_{i}=M_{\mathrm{p}}$, $10^{-2}M_{\mathrm{p}}$ and $10^{-4}M_{\mathrm{p}}$. 
Here, we take the expectation value of the PQ field during inflation to be 
$\left|\braket{\Phi}\right|=\left|\Phi\right|_{i}$ since the PQ field hardly move 
during inflation and it starts to oscillate soon after the end of inflation as described 
in Section~\ref{sec:The Dynamics of the Fields}. 
It is found that the axion can not be a main component of the dark matter 
because the constraint 
from the isocurvature perturbations is much stronger than that from the axion density 
in the chaotic inflation model. 
Moreover, the initial misalignment angle must be smaller than $O(10^{-2})$ 
so that the observational constraints are satisfied.
Here it should be noticed that the axion model with $N_{\mathrm{DW}}\ge 2$ can be excluded
by the isocurvature perturbation constraint 
if the PQ scalar settles at the potential minimum ($|\langle \Phi\rangle |=v$) during inflation.
Therefore, the present model succeeds in solving the serious isocuravture perturbation
problem in chaotic inflation for $|\Phi|_{i}\gg v$. 
As is seen from Eqs.\,\eqref{eq:axion abundance} and \eqref{eq:CDM isocurvature},
this result is almost unchanged even if $N_{\mathrm{DW}}$ is larger than two. 
In case of $N_{\mathrm{DW}}=6$, for instance, 
the upper bounds for $\theta_{a}$ and $v$ become about $1.2$ times larger
and about $0.5$ times smaller, respectively.

\subsubsection{$N_{\mathrm{DW}}=1$}
\label{subsec:N=1}

\begin{figure}[tbp]
\begin{tabular}{cc}
  \begin{minipage}{0.5\hsize}
  \begin{center}
     \includegraphics[width=7cm, clip]{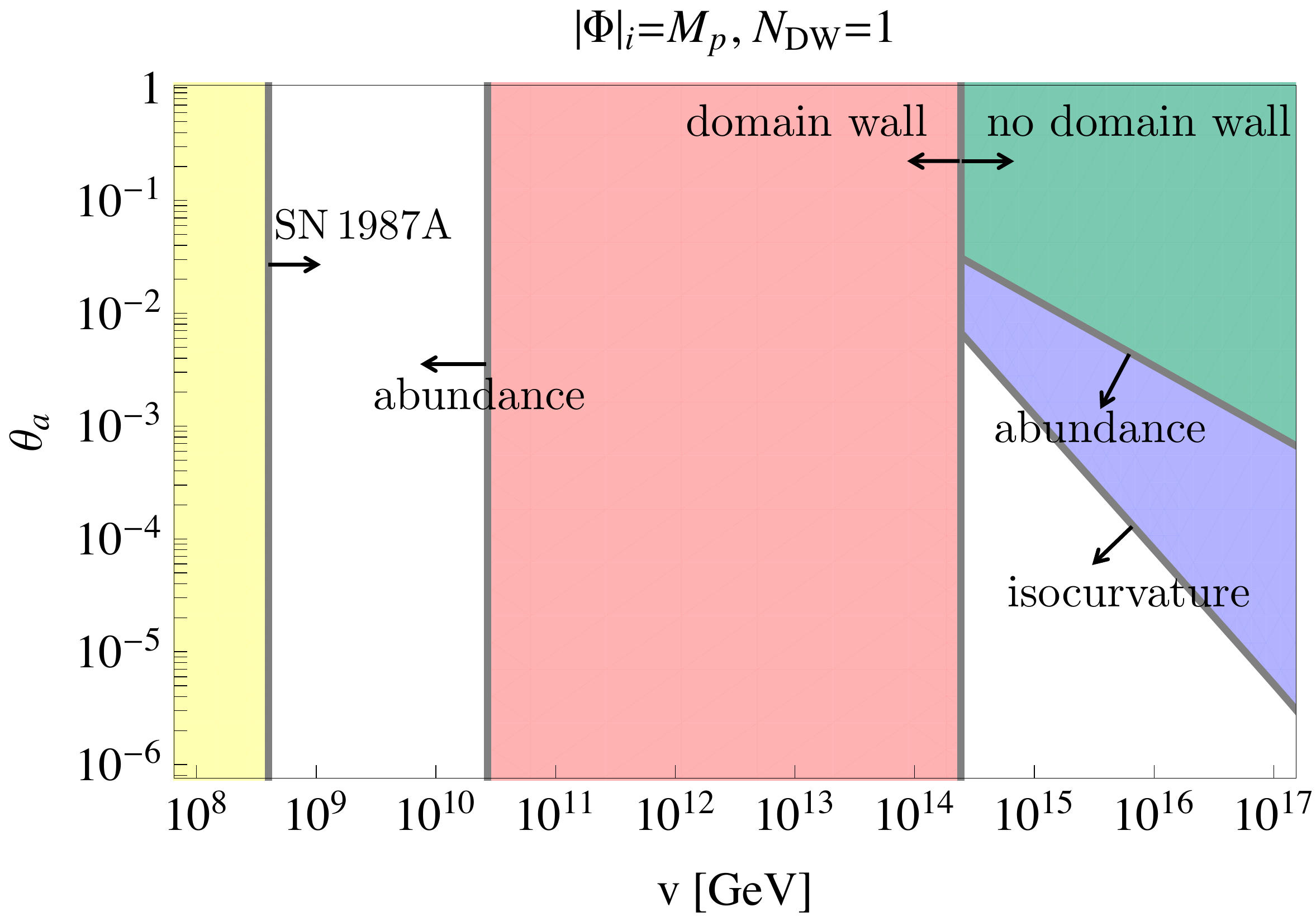}
  \end{center}
  \end{minipage}
  \begin{minipage}{0.5\hsize}
  \begin{center}
     \includegraphics[width=7cm, clip]{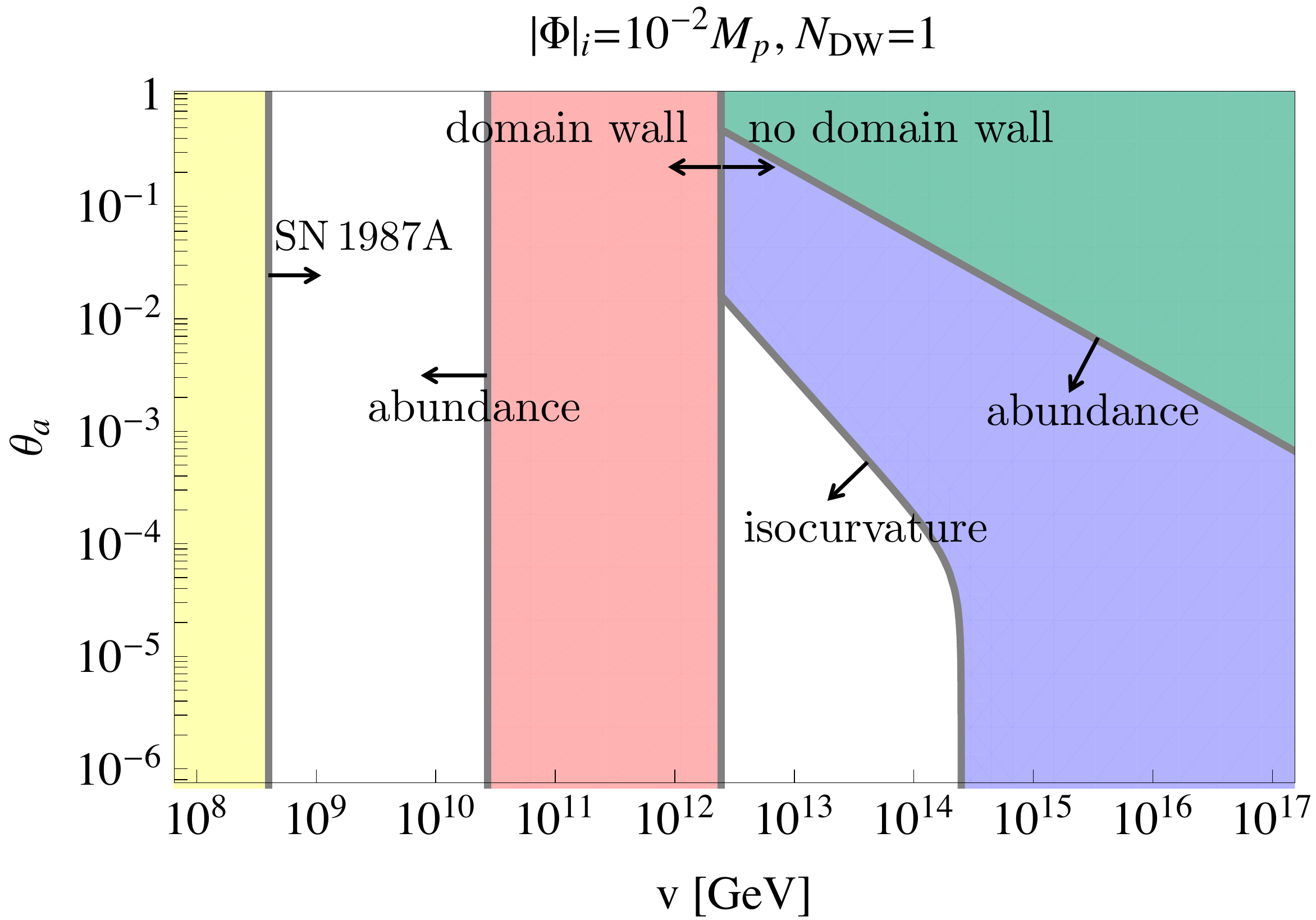}
  \end{center}
  \end{minipage}
\end{tabular}
  \begin{center}
    \includegraphics[width=7cm, clip]{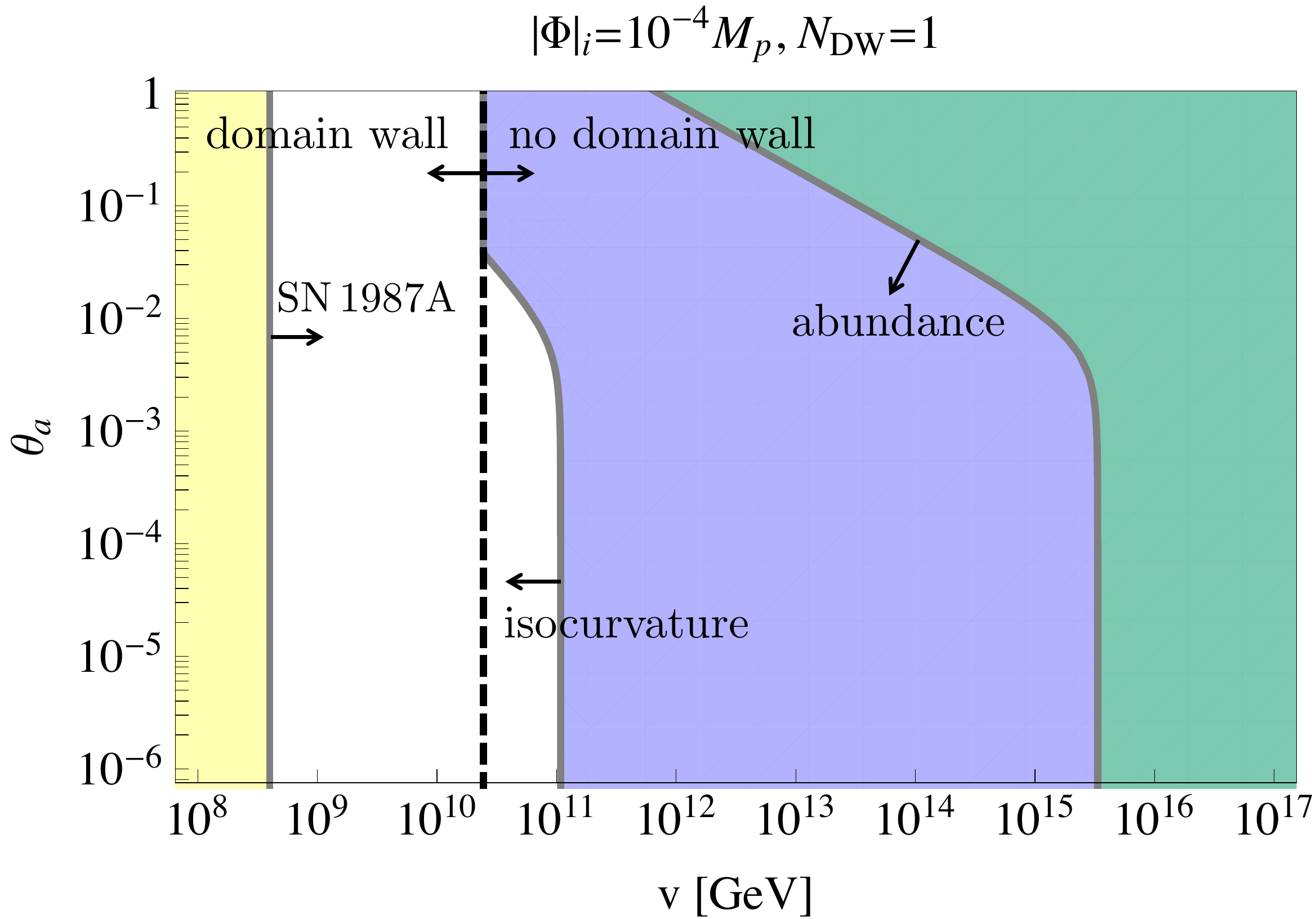}
  \end{center}
    \caption{Constraints on $v~(=N_{\mathrm{DW}}F_{a})$ and $\theta_{a}$ 
    for $N_{\mathrm{DW}}=1$ and various values of initial amplitude $\left|\Phi\right|_{\mathrm{i}}$ 
    in the chaotic inflation model, i.e.~$H_{\mathrm{inf}}=10^{13}\,\mathrm{GeV}$. 
    In the case of $v\gtrsim10^{-4}\left|\Phi\right|_{i}$, 
    the blue and the green regions are excluded by the observation of the isocurvature 
    perturbations and CDM abundance, respectively. 
    In the case of $v\lesssim10^{-4}\left|\Phi\right|_{i}$, 
    the red region is the constraint from the axion abundance and 
    the yellow region is excluded by the supernova 1987A.
    }
\label{fig:v-theta N=1}
\end{figure}

For $N_{\mathrm{DW}}=1$, the domain wall problem can be solved even if $U(1)_{\mathrm{PQ}}$ 
symmetry is broken after inflation as mentioned above. 
Therefore, we consider both regions $v\gtrsim10^{-4}\left|\Phi\right|_{i}$ and 
$v\lesssim10^{-4}\left|\Phi\right|_{i}$. 
In the former case, the expression for the present axion abundance Eq.\,\eqref{eq:axion abundance} 
and that for the CDM isocurvature perturbation Eq.\,\eqref{eq:CDM isocurvature} can be adopted 
since there is no symmetry restoration after inflation. 
On the other hand, the axion abundance comes from emission from string-wall system 
[Eq.\,\eqref{eq:axion abundance from string-wall}] 
in the latter case. 
Furthermore, there is another observational constraint coming from the cooling rate 
of supernova 1987A which imposes the lower limit on axion decay constant as 
$F_{a} \gtrsim4\times10^{8}\,\mathrm{GeV}$~\cite{Raffelt:2006cw}. 
Fig.\,\ref{fig:v-theta N=1} shows the parameter regions allowed by the result of 
the numerical simulations and the observational constraints for $N_{\mathrm{DW}}=1$ 
and three initial conditions $\left|\Phi\right|_{i}=M_{\mathrm{p}}$, $10^{-2}M_{\mathrm{p}}$ 
and $10^{-4}M_{\mathrm{p}}$. 
In the case of $v\gtrsim10^{-4}\left|\Phi\right|_{i}$, 
axion can not become a main component of dark matter and 
the initial misalignment angle must be smaller 
than $O(10^{-2})$ in the same way as $N_{\mathrm{DW}}\ge2$. 
In the case of $v\lesssim10^{-4}\left|\Phi\right|_{i}$, 
it is found that axion can become 
dark matter for $v\simeq3\times10^{10}~\mathrm{GeV}$ 
and $\left|\Phi\right|_{i}\gtrsim10^{-4}M_{\mathrm{p}}$.

\subsection{General inflation model}
\label{subsec:General inflation model}

\begin{figure}[tbp]
  \begin{tabular}{cc}
  \begin{minipage}{0.5\hsize}
  \begin{center}
    \includegraphics[width=6.5cm, clip]{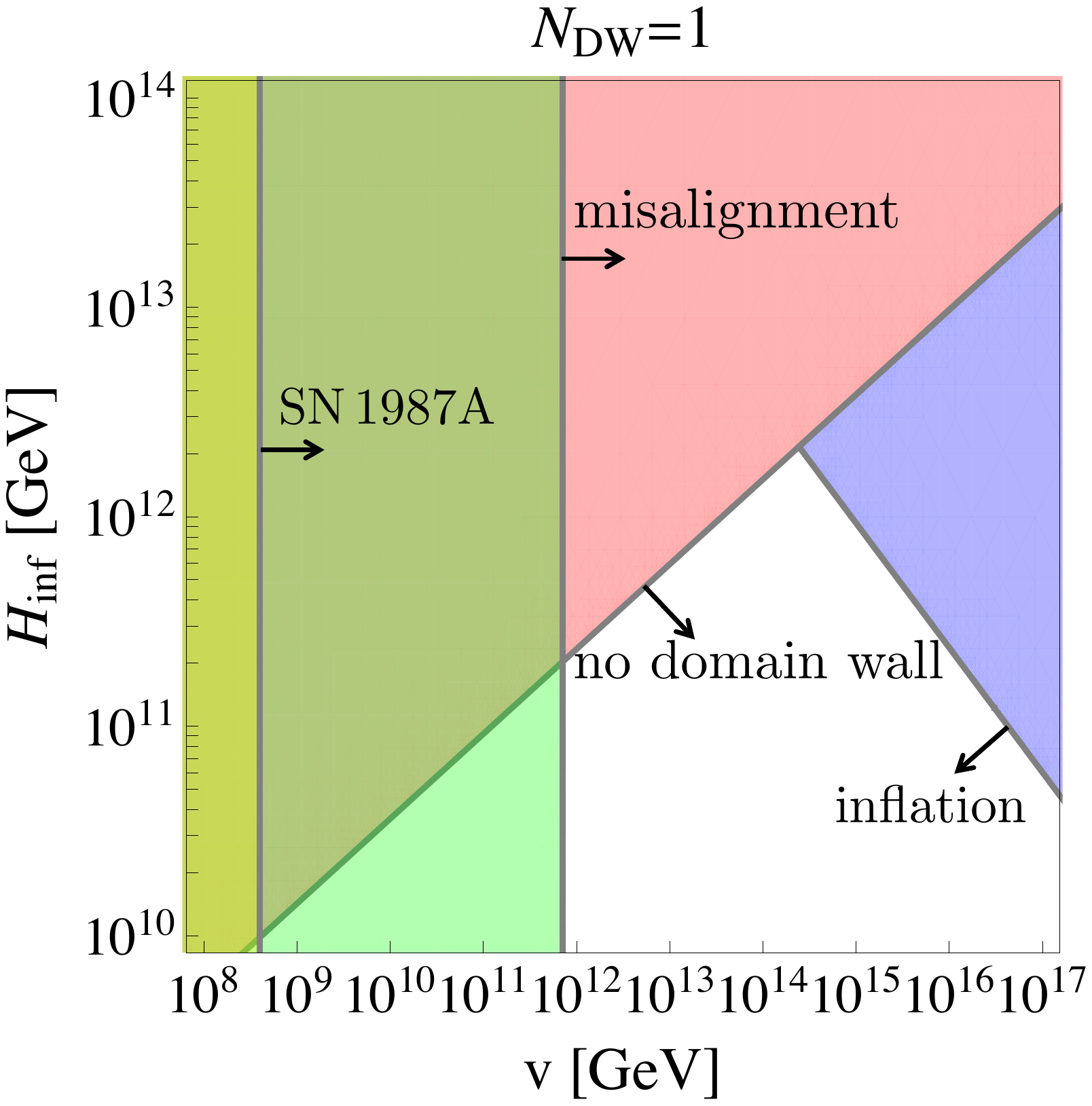}
  \end{center}
  \end{minipage}
  \begin{minipage}{0.5\hsize}
  \begin{center}
    \includegraphics[width=6.5cm, clip]{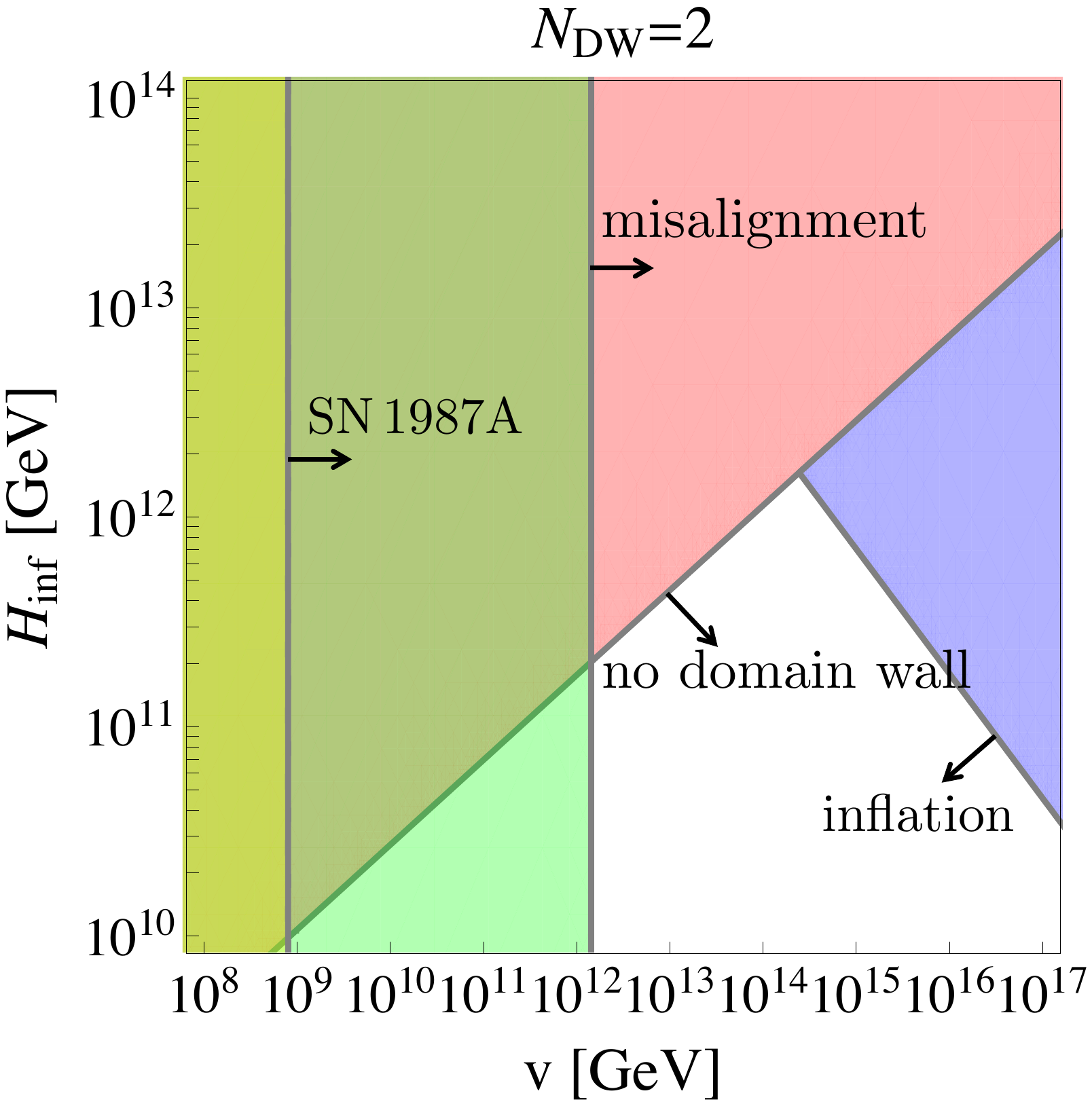}
  \end{center}
  \end{minipage}
  \end{tabular}
  \caption{The parameter regions for 
  $v~(=N_{\mathrm{DW}}F_{a})$ and $H_{\mathrm{inf}}$ allowed by some conditions and observational constraints 
  in case that axion is the main component of dark matter, 
  $\Omega_{a}h^2=\Omega_{\mathrm{CDM}}h^2$. 
  The yellow and the green regions represent the observational constraint 
  for the cooling rate of supernova 1987A 
  and the condition for the misalignment angle, 
  $\theta^2_{a}+\braket{\delta\theta^2_{a}}\lesssim1$, respectively. 
  The red region is excluded by the constraints 
  for the domain wall problem and CDM isocurvature perturbation.
  The blue region is the condition for the PQ field not to cause inflation.}
  \label{fig:v-H}
\end{figure}

So far we have assumed the chaotic inflation model  
and $H_{\mathrm{inf}}\simeq10^{13}~\mathrm{GeV}$. 
Now, we consider general models of inflation in which the Hubble parameter $H_{\mathrm{inf}}$ is smaller
and inflaton oscillates with the quadratic potential after the end of inflation. 
The value of the Hubble parameter determines only the initial amplitude 
of the fluctuations of the PQ field 
and the exponential growth of its fluctuations is independent of the Hubble parameter 
from our discussion in Section~\ref{sec:The Dynamics of the Fields}. 
Therefore, it is expected that the lower limit of the breaking scale $v$ which is necessary 
to avoid the domain wall problem is scarcely varied when the Hubble parameter changes 
by a few order. 
Indeed, performing some simulations for 
$H_{\mathrm{inf}}=10^{10}~\mathrm{GeV}$ 
and 
$\left|\Phi\right|_{i}=M_{\mathrm{p}}$,
we found that the PQ breaking scale must be larger than 
$(1-2)\times10^{14}~\mathrm{GeV}$.
This is consistent with the above result in the chaotic inflation model in Section~\ref{sec:Numerical Simulations}.

If we impose a condition that axions are the main component of the dark matter 
$\Omega_{a}h^2=\Omega_{\mathrm{CDM}}h^2$, 
the power spectrum of the isocurvature perturbations 
Eq.\,\eqref{eq:CDM isocurvature} is a function of the Hubble parameter $H_{\mathrm{inf}}$, 
the initial value of the PQ field $\left|\Phi\right|_{i}$, 
the breaking scale of PQ symmetry $v$ 
and the domain wall number $N_{\mathrm{DW}}$. 
Avoiding the domain wall problem constrains the initial value $\left|\Phi\right|_{i}$
to satisfy Eq.\,\eqref{eq:constraint for v}.
In  addition,  in order for the PQ field not to cause inflation, 
$\left|\Phi\right|_{i}\lesssim M_{\mathrm{p}}$ is should be satisfied. 
Since the amplitude of the isocurvature perturbations has a minimum value 
when $\left|\Phi\right|_{i}$ has the maximum value,
it is needed that the power spectrum of the isocurvature perturbations 
for $\left|\Phi\right|_{i}=10^{4}v$ and $\left|\Phi\right|_{i}=M_{\mathrm{p}}$ satisfies 
the observational constraint simultaneously as
\begin{equation}
   \left.\mathcal{P}_{\mathcal{S}_{\mathrm{CDM}}}(k_{0})\right|_{\left|\Phi\right|_{i}=10^{4}v} <
   \frac{\beta_{\mathrm{iso}}}{1-\beta_{\mathrm{iso}}}\mathcal{P}_{\zeta}(k_{0}),
   ~~~~\left.\mathcal{P}_{\mathcal{S}_	{\mathrm{CDM}}}(k_{0})\right|_{\left|\Phi\right|_{i}=M_{\mathrm{p}}} <
   \frac{\beta_{\mathrm{iso}}}{1-\beta_{\mathrm{iso}}}\mathcal{P}_{\zeta}(k_{0}).
   \label{eq:CDM isocurvature for v-H}
\end{equation}
Furthermore, there is a condition that the misalignment angle is less than unity, 
$\theta^2_{a}+\braket{\delta\theta^2_{a}}\lesssim1$, by definition.
Fig.~\ref{fig:v-H} shows the constraints Eq.\,\eqref{eq:CDM isocurvature for v-H} for $\Omega_{a}h^2=\Omega_{\mathrm{CDM}}h^2$, 
the lower bound of the breaking scale $v$ 
from the observation of SN 1987A 
and the condition for the misalignment angle. 
From the figure, it is found that axion can be the main component of the dark matter 
avoiding both domain wall and isocurvautre perturbation problems
for $v\simeq10^{12-16}\,\mathrm{GeV}$
and $H_{\mathrm{inf}}\lesssim 10^{11-12}\,\mathrm{GeV}$ 
in general inflation models. 
It should be noticed that 
the present argument does not apply 
to the string axion model, i.e. $F_{a}\simeq10^{16}\,\mathrm{GeV}$~\cite{Kawasaki:1997ct,Svrcek:2006yi},
because the dynamics of PQ field is different from that of our model. 
In practice, for the string axion model
the Hubble parameter during inflation should be less than $10^{9}\,\mathrm{GeV}$ 
in order to avoid the isocurvature problem (e.g. see~\cite{Hikage:2012be}).

\section{Conclusion}
\label{sec:Conclusion}

We have considered the axion model in which $U(1)_{\mathrm{PQ}}$ symmetry is spontaneously broken
during inflation and the value of the PQ field is large enough to suppress 
the axion isocurvature perturbations. 
The homogeneous part of the PQ field oscillates along its potential in the matter dominated 
universe after inflation and their fluctuations grows exponentially due to 
the parametric resonance through the self-coupling of the PQ field.
If the fluctuations is large, the $U(1)_{\mathrm{PQ}}$ symmetry is restored and 
many stable strings are formed, which results in the domain wall problem. 
Calculating the formation of stable axionic strings using lattice simulations, 
we have found that $U(1)_{\mathrm{PQ}}$ breaking scale should satisfy 
$v\gtrsim10^{-4}\left|\Phi\right|_{i}$, where $\left|\Phi\right|_{i}$ is the value of PQ field 
at the beginning of oscillation, for the stable strings not to be formed. 
This result is much weaker than that in Ref.~\cite{Kasuya:1999hy} because  it assumed
the radiation dominated universe where the parametric resonance is more significant
as shown in Section~\ref{sec:The Dynamics of the Fields}.
Combining our numerical result with the observational constraints from the matter density of the universe 
and the CDM isocurvature perturbations, it is found that the axion model is consistent with
chaotic inflation if the initial misalignment angle $\theta_a$ is less than $O(10^{-2})$.
In this case axions can not be the main component 
of dark matter in the chaotic inflation model with Hubble parameter
$H_{\mathrm{inf}}\simeq10^{13}\,\mathrm{GeV}$. 
However, when we consider general inflation models 
axion can account for the dark matter without 
the domain wall problem nor isocurvature perturbation problem 
for $v\simeq10^{12-16}\,\mathrm{GeV}$ and $H_{\mathrm{inf}}\lesssim10^{12}\,\mathrm{GeV}$. 
This implies that topological inflation model~\cite{Vilenkin:1994pv,Izawa:1998rh,Kawasaki:2000tv} 
is marginally allowed since the Hubble parameter during inflation 
is estimated as $H_{\mathrm{inf}}\simeq10^{12}\,\mathrm{GeV}$\cite{Harigaya:2012hn}.

\section*{Acknowledgements}

We thank Ken'ichi Saikawa for useful discussions. 
This work is supported by Grant-in-Aid for Scientific research from
the Ministry of Education, Science, Sports, and Culture (MEXT), Japan,
No.\ 25400248 (M.K.), No.\ 21111006 (M.K.), No.\ 22244021 (T.T.Y.) and also 
by World Premier International Research Center
Initiative (WPI Initiative), MEXT, Japan. 


\end{document}